\documentclass[a4paper]{article}

\usepackage{INTERSPEECH2021}
\usepackage{graphicx, subfig}
\usepackage{xcolor,stfloats}
\usepackage{latexsym,bm}        
\usepackage{amsmath,amssymb} 
\usepackage{bbding}
\usepackage{multirow}
\title{Explore Long-Range Context feature for Speaker Verification}
\name{Zhuo Li, Runqiu Xiao, Xiaoxiao Miao, Wenchao Wang{\rm\textsuperscript{†}}, Pengyuan Zhang\thanks{\textsuperscript{†}Corresponding author.}}
\address{
 Key Laboratory of Speech Acoustics and Content Understanding, Institute of Acoustics, Chinese Academy of Sciences, Beijing, China\\
 University of Chinese Academy of Sciences, Beijing, China
}
\email{\{lizhuo,xiaorunqiu,miaoxiaoxiao,wangwenchao,zhangpengyuan\}@hccl.ioa.ac.cn}
\begin{document}

\maketitle
\begin{abstract}
Capturing long-range dependency and modeling long temporal contexts is proven to benefit speaker verification tasks. In this paper, we propose the combination of the Hierarchical-Split block(HS-block) and the Depthwise Separable Self-Attention(DSSA) module to capture richer multi-range context speaker features from a local and global perspective respectively.
Specifically, the HS-block splits the feature map and filters into several groups and stacks them in one block, which enlarges the receptive fields(RFs) locally. 
The DSSA module improves the multi-head self-attention mechanism by the depthwise-separable strategy and explicit sparse attention strategy to model the pairwise relations globally and captures effective long-range dependencies in each channel. Experiments are conducted on the Voxceleb and SITW. Our best system achieves 1.27\% EER on the Voxceleb1 test set and 1.56\% on SITW by applying the combination of HS-block and DSSA module.


%
\end{abstract}

\section{Introduction}

Speaker verification(SV) has developed rapidly over the years. End-to-end SV systems have emerged in recent years and achieved the state-of-the-art performance. X-vector\cite{xvectorformal} is one of the most popular end-to-end SV systems. X-vector use the Time Delay Neural Network (TDNN) as a feature extractor to generate frame-wise speaker embedding. Afterward, a statistic pooling is bulit on top to produce the fixed-dimensional utterance-level speaker embedding. This embedding are further processed by fully-connected layers and an output layer. Finally, a loss function is used to optimize the entire network in an end-to-end manner. Although X-vector introduces TDNN and statistic pooling to consider the dependencies of contiguous frames, it is still challenging to model long-range dependencies due to the limited RFs and slightly weak deep representation ability.

With the rising popularity of the x-vector, the topology of the network has achieved significant architectural improvements. The popular ResNet\cite{resnet} architectures are introduced into the speaker verification tasks\cite{resnet00,resnet01,resnet02} as the feature extractor to generate frame-level representations, which achieve significant improvement. Despite this, it's still lacking in capturing long-term dependencies. Prior work\cite{RFs} has shown that the empirical RFs gained by a chain of convolutions is much smaller than the theoretical RFs by experiments, especially in deeper layers. Another work\cite{efrfs} has proven that the distribution of impact within an effective receptive field is limited to a local region and converged to the gaussian.
Therefore, whether TDNN or CNN model, capturing long-range dependency is insufficient.

To address these problems, studies\cite{lstm0,lstm1,lstm2,lstm3} insert the LSTM on the top of or inside the backbone network for SV tasks, because the LSTM is good at modeling the long-term information. The authors in \cite{lstm4} combine the BLSTM and ResNet into one unified architecture. Also, there are numerous approaches to capture long-range dependencies in computer vision field. These could be roughly divided into two categories\cite{two}, local-based approaches and global-based approaches. The former is to increase the local RFs by dilated operation\cite{dilated} or split-stack operation\cite{gao2019res2net}, the latter is to extract long-range information by modeling the pairwise relations globally based on the attention mechanism\cite{attention}.

Inspired by study\cite{two}, we explore long-range context feature for SV from local and global  perspectives.
From the local perspective, we introduce a novel Hierarchical-Split block\cite{hsresnet} to enlarge the RFs and generate multi-scale feature representations, named HS-ResNet\cite{hsresnet}.
From the global perspective, we improve the attention mechanism to model long-term dependencies and learn a rich hierarchy of associative features across long-time duration.
Besides, to force the network to focus on the most relevant segments, the explicit sparse attention mechanism\cite{sparse} is explored. 


Our contributions are summarized as follows.
\begin{itemize}
\vspace{-0.1mm}
\itemsep=0pt
\item we introduce a novel Hierarchical-Split block to enlarge the RFs in a single block for SV tasks, named HS-resnet.
\item we propose an innovative plug-and-play module based on the attention mechanism, DSSA module. DSSA module is flexible and extendable, and it can easily be plugged into multiple mature architectures to improve performance.
\item To improve the concentration of attention on the global context and avoid the effect of irrelevant information, we explore the sparse attention mechanism in this paper.

\vspace{-0.1mm}
\end{itemize} 
The organization of this paper is as follows. The proposed HS-block is described detailedly in Section 2. Section 3 demonstrates the DSSA module, which can be taken as a plug-and-play module based on the attention mechanism. The experiment settings and results are given in Section 4. Section 5 concludes the paper.

%
%

\begin{figure*}\vspace{-0.3cm}
\centering
\includegraphics[width=1.0\linewidth]{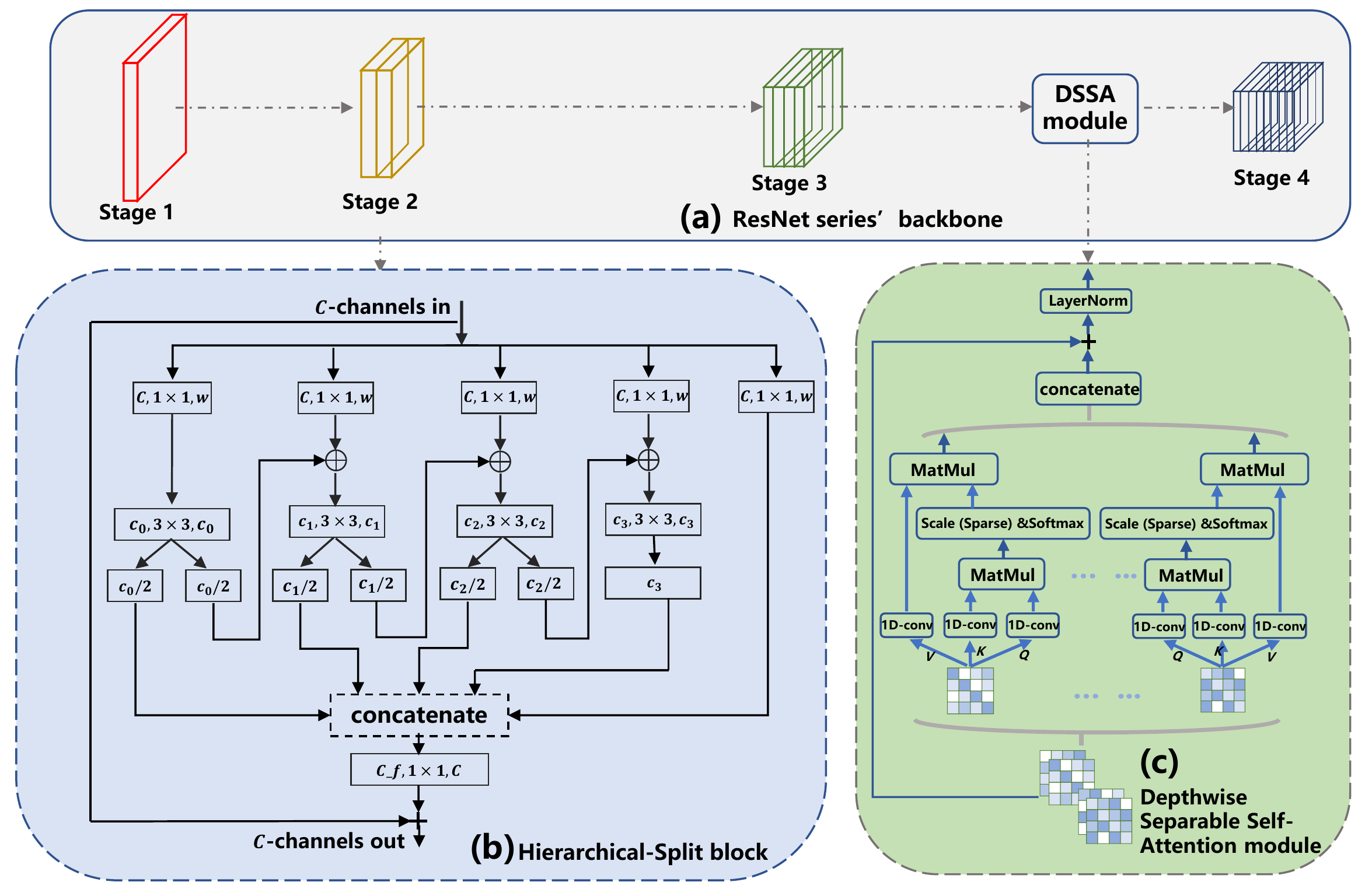}
\centering
\caption{ detail structure of  HS-ResNet50 and Depthwise Separable Self-Attention module for the combined system} 
\label{gene}
\vspace{-0.5cm}
\end{figure*}

\section{Local-based Approach}
Local-based approaches enlarge the local receptive field through pooling, dilated, \emph{split-stack} and other operations usually. Hierarchical-Split block is one of the approaches by \emph{split-stack} operations.
The structure of the HS-block is depicted in Figure\ref{gene}(b). The $3*3$ block in the typical ResNet is modified. After the $1*1$ convolution, the feature maps are split equally into \emph{s} groups, denoted by $x_i$, the $3*3$ convolution filters are also replaced by several groups, denoted as $\mathcal{F}_i\left( \right)$. Each $x_i$ will be fed into $\mathcal{F}_i\left( \right)$ and the output feature maps are denoted by $y_i$.  Here, each $y_i$ is split into two sub-groups, denoted $y_{i,1}$ and $y_{i,2}$. Then, $y_{i,2}$ is concatenated with the following group $x_{i+1}$, and then sent into  $\mathcal{F}_{i+1}\left( \right)$. All $y_{i,1}$ are concatenated in the channel dimension as the output of the $3*3$ convolution filters, denoted as $y_i$. Especially, each $y_{i,1}$ has different channels and RFs, and more channels $y_{i,1}$ contains, the larger RFs are gained. In this manner, the feature maps could contain detailed information and larger-range dependencies. $\bigoplus$ means two feature maps are concatenated in the channel dimension. 
\begin{equation}
\setlength\abovedisplayskip{2mm}
\setlength\belowdisplayskip{1mm}
y_i=\left\{
\begin{array}{rcl}
x_i, & & {i = 1}\\
\mathcal{F}_{i}(x_{i} \bigoplus y_{i-1,2}), & & { 1< i <= s}\\
\end{array} \right.
\label{equ1}
\end{equation}
Besides, the hyperparameters $s$ and $t$ are used to control the HS-ResNet's parameters and complexity. $s$ means groups the feature maps are divided into, $t$ means how many times the number of channels will be expanded. $k$ means the size of convolutional kernel, $w$ means the number of channels. The computational complexity of HS-block can be calculated as follows. 
\vspace{-1mm}
\begin{eqnarray}
\setlength\abovedisplayskip{0mm}
\setlength\belowdisplayskip{0mm}
C_0&=&w\times \frac{t}{s} \quad C_i = C_0 + \frac{C_{i-1}}{2} \\
\emph{PARAM}&=& k^2\times(\sum_{i=0}^{s-2} C_i^2 + C_0^2) \\
&=&C_0^2\times(4s-\frac{29}{3}+16\times 2^{-s} -\frac{16}{3}\times 2^{-2s})\nonumber
\end{eqnarray}
\vspace{-6.6mm}

\section{Global-based Approach}
Global-based approaches are generally based on the attention mechanism because they can model the pairwise relations from the global perspective.
\subsection{Multi-Head Scaled Dot-Product Attention}
Multi-head Scaled Dot-Product Attention, a core module of the famous Transformer architecture, is also the core of our module. Multi-head attention consists of multiple Scaled Dot-Product Attention, and each Scaled Dot-Product Attention is calculated independently. Each Scaled Dot-Product Attention consists of three inputs: queries, keys and values. Each query's output will be calculated the dot products with all keys' outputs, the results are divided by $\sqrt{d_k}$ and applied a softmax function to obtain the weights of the values' output. In practice, the above calculations are implemented in the form of matrices. The outputs are computed as:
\begin{align}
&Attention(Q,K,V) = softmax(QK^T/\sqrt{d_k})V
\end{align}
The Multi-Head Scaled Dot-Product Attention's output is obtained by concatenating each Scaled Dot-Product Attention's output and feeding them into another linear projection. The formula is as follows:
\begin{flalign}
&MultiHead(Q,K,V) = Concat(head_1,...,head_h)W^O \nonumber\\
&where \quad head_i = Attention(QW_i^Q,KW_i^K,VW_i^V)
\end{flalign}
In the equation, $d_q=d_k=d_v=d_{model}/h=d$, the projection matrices ${W_i^Q,W_i^K,W_i^V} \in \mathbb{R}^{d_{model}\times d}$, $W^O \in \mathbb{R}^{hd_k\times d_{model} }$. 
\subsection{Depthwise Separable Self-Attention Module}
Here are a few challenges when using Multi-Head Scaled Dot-Product Attention in CNN structures, especially the 
parameters. The depthwise separable convolution strategy, which decomposes ordinary convolutions into depthwise convolution and pointwise convolution, is introduced in the module to avoid the massive parameters.

In this DSSA module, the depthwise separable convolution applies a single Scaled Dot-Product Attention to each input channel. It is worth noting that the pointwise convolution is abandoned. The reason is explained in the next section. As decipted in Figure\ref{gene}(c), the input feature maps are split into several groups, and the number of groups is equal to the number of channels in the feature maps. That is, each group only contains one channel of the feature maps. 
Then, each group is fed into three independent 1D-convolution to generate the queries, keys and values. The dimension of these is the same as the input.
Specifically, the DSSA module takes as input a $C\times T \times W$ feature map and produces three independent $C\times T \times W$ feature maps after the convolution layers. After that, the dot products between each query and each key are calculated and divided by $\sqrt{d_k}$. Before applying a softmax function to obtain the weights on the values', another square operation acts on the weights. This operation is to maintain weights in the normal range and avoid it too big or too small. On each group of output values, we perform the layer-norm operation in parallel. Each group of output values are concatenated and performed the layer-norm operation as the final output.
\begin{align}
P_c & = \sqrt{Q_cK_c^T/\sqrt{d_k}} \\
C &= Concat(Softmax(P_c)V_c) \\
y & = LayerNorm(x + C))
\end{align}

\subsection{Sparse Self-Attention mechanism}
In addition, to explore the effect of critical frames, the explicit sparse attention mechanism is designed in this module as an option. The explicit sparse attention only pays attention to the k most contributive states. 
The attention weights are degenerated to the sparse attention through top-k selection. The $k$ largest elements of each row in the attention weight matrix are selected, and the others are replaced with $-\infty$,

\begin{equation}
\setlength\abovedisplayskip{1mm}
\setlength\belowdisplayskip{1mm}
\mathcal{M}(P_c,k)_{ij}=\left\{
\begin{array}{rl}
P_{c,{ij}},&if \ P_{c,{ij}} \geq t_i \\
-\infty,& if \ P_{c,{ij}}< t_i
\end{array} \right.
\end{equation}
where $t_i$ means the $k$-th largest value of row $i$.
\section{Experiments and Results}
\subsection{Datasets}
We conduct experiments on the following datasets, VoxCeleb1(Vox1)~\cite{vox1}, VoxCeleb2(Vox2)~\cite{vox2} and The Speakers in the Wild (SITW)~\cite{sitw}. 
The \emph{dev} part of Vox1\&2 are used as the training part respectively, which contains 1211 speakers and 5994 speakers separately. Data augmentation is not used in all experiments.
The Vox1 test sets, Vox1-E(cleaned), Vox1-H(cleaned) and \emph{core-core} trials in the SITW database are used as evaluation sets.
The VoxCeleb1 test sets is used in the analysis part mainly, while the VoxCeleb1-E/H(cleaned) and \emph{core-core} trials of SITW datasets are used to prove the generalisability and robustness of our model in the final.

\subsection{Settings}
\textbf{Data preprocessing.} We use 64-dimensions FBanks as the raw acoustic features, which extracted from 25ms frames with 10ms overlap, spanning the frequency range 0-8000Hz. No voice activity detection (VAD) is applied. 

\textbf{Model.} Following the previous work in \cite{resnet00}, the standard ResNet-34 and ResNet-50 architecture are used in our experiments. The initial number of channels is set as 16 when the training data is VoxCeleb 1 and set as 32 when training data is VoxCeleb2. Only the mean of the frame-level features is used. To maintain the similar number of parameters with ResNet50, $t$ is set as 1.5 when $s$ is set as 8 in the following experiments according to Equation(3) in the HS-ResNet. 
Besides, experiments show that inserting the DSSA module between stages 3 and 4 achieves better improvements. Thus, for convenience, the DSSA module is applied between stages 3and 4 if not special specified.

\textbf{Training and Testing.}
In the training stage, mini-batch size of 32 is used to train models in all experiments.
Softmax with cross entropy loss is used to train our model. Stochastic gradient descent (SGD) with momentum 0.9, weight decay 1e-3 is utilized. The learning rate is set to 0.1, 0.01, 0.001 and is switched when the training loss plateaus. Each speech sample in the training stage is sampled for L frames from each speech sample. The chunk-size L is randomly sampled from the interval [L1;L2], and the interval is set to [200,400], [300,500] and [400,600] in the three training stages.
In the testing stage, cosine similarity is applied as the back-end scoring method. The performance of different systems is gauged in terms of the EER and minDCF(0.01).

\subsection{Analysis and Results}
\subsubsection{Local-based Approach: HS-ResNet}
\label{4.3.1}

As displayed in Table\ref{vox1}, HS-ResNet50 exceeds the ResNet50 by 18\% on EER and 15\% on minDCF. The stronger ability of modeling long-range dependencies with HS-ResNet are proven that it is able to achieve great performance improvements by experiments. Its unique \emph{split-stack} structure is effective to collect more long-scale features and more long-range imformation. 

\begin{table}[htbp]
  \centering
  \caption{The results on Vox1-Test when the training data is Vox1}
    \begin{tabular}{ccc}
    \toprule
    System & EER(\%) & MinDCF(0.01) \\
    \midrule
    ResNet34 & 3.807 & 0.3465 \\
    \midrule
    ResNet34+DSSA & 3.123 & 0.296 \\
    \midrule
    ResNet50 & 3.966 & 0.3755 \\
    \midrule
    ResNet50+DSSA & 3.197 & 0.3054 \\
    \midrule
    HS-ResNet50 & 3.203 & 0.3144 \\
    \midrule
    HS-ResNet50 + DSSA & 2.741 & 0.2789 \\
    \bottomrule
    \end{tabular}%
  \label{vox1}%
\end{table}%
%

\subsubsection{Global-based Approaches: The Depthwise Separable Self-Attention Module}
\label{4.3.2}

The performance of systems with the DSSA module is shown in Table \ref{vox1}. They all achieve  great performance improvements. ResNet34+DSSA system achieves 18\% improvements on  EER and 15\% on minDCF when the training data is Vox1.
%
%
%

\begin{table*}[htbp]
\vspace{-0.8cm}
  \centering
  \caption{EER and MinDCF performance of all systems on the standard Vox1 test set when the training data is Vox2}
	\resizebox{\textwidth}{22mm}{
    \begin{tabular}{ccccccccc}
    \toprule
    \multirow{2}[4]{*}{System} & \multicolumn{2}{c}{Vox1-Test} & \multicolumn{2}{c}{Vox1-E} & \multicolumn{2}{c}{Vox1-H} & \multicolumn{2}{c}{SITW-c-c} \\
\cmidrule{2-9}          & EER(\%) & MinDCF(0.01) & EER(\%) & MinDCF(0.01) & EER(\%) & MinDCF(0.01) & EER(\%) & MinDCF(0.01) \\
    \midrule
    ResNet34 & 1.914 & 0.194 & 1.787 & 0.2024 & 3.231 & 0.3061 & 3.253 & 0.312 \\
    \midrule
    ResNet34+DSSA & 1.617 & 0.1385 & 1.503 & 0.176 & 2.763 & 0.2648 & 2.05  & 0.1851 \\
    \midrule
    ResNet50 & 2.031 & 0.1957 & 1.857 & 0.2088 & 3.279 & 0.3187 & 3.417 & 0.3166 \\
    \midrule
    ResNet50+DSSA & 1.76  & 0.1566 & 1.548 & 0.1731 & 2.839 & 0.2643 & 2.378 & 0.2124 \\
    \midrule
    HS-ResNet50 & 1.458 & 0.1356 & 1.341 & 0.1631 & 2.581 & 0.243 & 2.734 & 0.2554 \\
    \midrule
    HS-ResNet50+DSSA & 1.273 & 0.1005 & 1.19  & 0.1469 & 2.292 & 0.2188 & 1.558 & 0.1564 \\
    \bottomrule
    \end{tabular}}
  \label{all}%
\vspace{-0.3cm}
\end{table*}%

\subsubsection{Global-based Approaches: Sparse Self-Attention Mechanism}
Although systems with DSSA module have obtained great improvements, the weights in the attention weights matrix vary close to 0 and are a little similar to each other after analyzing the weight matrix. Therefore, we try to apply sparse attention weight matrix to make the network focus on the most critical frames. After experiments, it's found that the phenomenon is alleviated, but performance has slightly decreased. The results are shown in Table\ref{topk}. The phenomenon will be analyzed and explored in our future work.

\begin{table}[htbp]
\vspace{-0.2cm}
  \centering
  \caption{Exploration of the effect of differnet $k$}
\vspace{-0.3cm}
    \begin{tabular}{cccc}
    \toprule
    System & top-k & EER(\%) & MinDCF(0.01) \\
    \midrule
    \multirow{2}[4]{*}{ResNet34\_DSSA} & 10    &  3.277     & 0.3081  \\
\cmidrule{2-4}          & 40    & 3.388       & 0.3196 \\
    \bottomrule
    \end{tabular}%
  \label{topk}%
\end{table}%

\subsubsection{ Complementary Analysis and Generalizability Analysis}
\label{4.3.3}
After the above experiments, we wonder that if these two approaches are complementary.
The local-based approach focuses on enlarging the local receptive field by stacking convolution filters in a single layer. The global-based approach focuses on implementing pairwise entity interactions with a content-based addressing mechanism to integrate information and extract long-range context features. In theory, they are complementary to each other. Thus, we combined these two approaches into one structure. The results are shown in Table\ref{vox1}\&\ref{all}. we combine the HS-block and DSSA module into one unified architecture and yield further advantage. The result suggests great complementarity to local-based approach and global-based approach and offers a reference for modeling long-range dependencies.

\begin{figure}[hb]
\vspace{-0.1cm}
\centering
\includegraphics[width=1.0\linewidth]{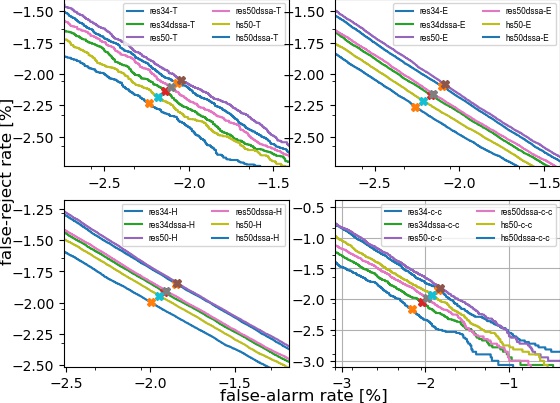}
\centering
\vspace{-0.6cm}
\caption{ DET curves for all systems} 
\label{det}
\vspace{-0.6cm}
\end{figure}
To prove the generalisability of the proposed method, the Vox1-E\&H and SITW \emph{c-c} are used to eval the performance. Experiments results are shown in Table\ref{all}. Combining two approaches achieve similar performance improvement on Vox1-E\&H trials as Vox1-Test, about 30\% on EER and minDCF, but they achieve over 50\% improvement on SITW \emph{c-c}. We attribute this great improvement to \emph{c-c}'s longer utterance than Vox1.

\subsubsection{ Params and Inference time Analysis}
\begin{figure}[hb]
\vspace{-0.2cm}
\centering
\includegraphics[width=1.0\linewidth]{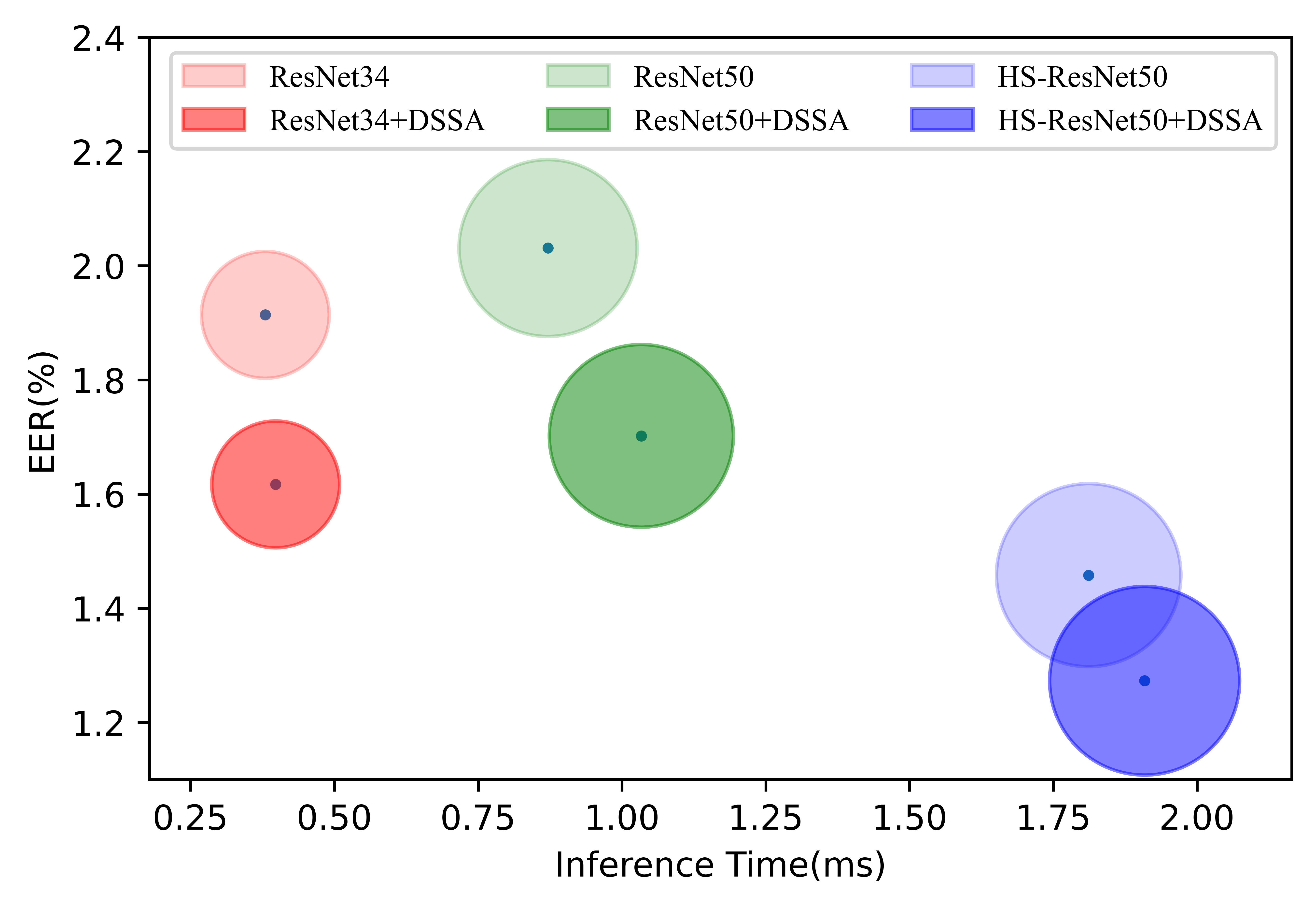}
\centering
\vspace{-0.6cm}
\caption{ Comparing the EER-inference time of all systems, the circle area reflects the params of system} 
\label{inf}
\vspace{-0.3cm}
\end{figure}

Finally, the number of parameters and the inference time of all systems are shown in Figure\ref{inf}. Although the number of parameters does not increase too much in HS-ResNet, the inference time increases by 80\% over ResNet50, because calculating the outputs of stacked filter groups in one single layer is time-consuming. While, the DSSA module not only increase too much parameters, but also maintain the inference time almostly. 
\section{Conclusions}

In this paper, we propose two approaches to capture long-range dependencies and improve performance for speaker verification. The first approach replaces the $3*3$ convolution with HS-block to enlarge the local RFs. The second proposes the DSSA module, which integrates information from the global perspective in each channel's of feature maps. Experiments show that both of these approaches yield performance improvements and the DSSA module is flexible and extendable. Besides, combining these approaches into one structure obtain further improvements, which offers an useful reference for further research. 


\bibliographystyle{IEEEtran}

\bibliography{mybib}

\begin{thebibliography}{10}
\providecommand{\url}[1]{#1}
\csname url@samestyle\endcsname
\providecommand{\newblock}{\relax}
\providecommand{\bibinfo}[2]{#2}
\providecommand{\BIBentrySTDinterwordspacing}{\spaceskip=0pt\relax}
\providecommand{\BIBentryALTinterwordstretchfactor}{4}
\providecommand{\BIBentryALTinterwordspacing}{\spaceskip=\fontdimen2\font plus
\BIBentryALTinterwordstretchfactor\fontdimen3\font minus
  \fontdimen4\font\relax}
\providecommand{\BIBforeignlanguage}[2]{{%
\expandafter\ifx\csname l@#1\endcsname\relax
\typeout{** WARNING: IEEEtran.bst: No hyphenation pattern has been}%
\typeout{** loaded for the language `#1'. Using the pattern for}%
\typeout{** the default language instead.}%
\else
\language=\csname l@#1\endcsname
\fi
#2}}
\providecommand{\BIBdecl}{\relax}
\BIBdecl

\bibitem{xvectorformal}
D.~Snyder, D.~Garcia-Romero, G.~Sell, D.~Povey, and S.~Khudanpur, ``X-vectors:
  Robust dnn embeddings for speaker recognition,'' in \emph{2018 IEEE
  International Conference on Acoustics, Speech and Signal Processing
  (ICASSP)}.\hskip 1em plus 0.5em minus 0.4em\relax IEEE, 2018, pp. 5329--5333.

\bibitem{resnet}
K.~{He}, X.~{Zhang}, S.~{Ren}, and J.~{Sun}, ``Deep residual learning for image
  recognition,'' in \emph{2016 IEEE Conference on Computer Vision and Pattern
  Recognition (CVPR)}, 2016, pp. 770--778.

\bibitem{resnet00}
W.~Cai, J.~Chen, and M.~Li, ``Exploring the encoding layer and loss function in
  end-to-end speaker and language recognition system,'' 04 2018.

\bibitem{resnet01}
\BIBentryALTinterwordspacing
C.~Li, X.~Ma, B.~Jiang, X.~Li, X.~Zhang, X.~Liu, Y.~Cao, A.~Kannan, and Z.~Zhu,
  ``Deep speaker: an end-to-end neural speaker embedding system,'' \emph{CoRR},
  vol. abs/1705.02304, 2017. [Online]. Available:
  \url{http://arxiv.org/abs/1705.02304}
\BIBentrySTDinterwordspacing

\bibitem{resnet02}
N.~Li, D.~Tuo, D.~Su, Z.~Li, and D.~Yu, ``Deep discriminative embeddings for
  duration robust speaker verification,'' 09 2018, pp. 2262--2266.

\bibitem{RFs}
B.~Zhou, A.~Khosla, A.~Lapedriza, A.~Oliva, and A.~Torralba, ``Object detectors
  emerge in deep scene cnns,'' 12 2014.

\bibitem{efrfs}
W.~Luo, Y.~Li, R.~Urtasun, and R.~Zemel, ``Understanding the effective
  receptive field in deep convolutional neural networks,'' \emph{arXiv preprint
  arXiv:1701.04128}, 2017.

\bibitem{lstm0}
\BIBentryALTinterwordspacing
X.~Miao, I.~McLoughlin, and Y.~Yan, ``{A New Time-Frequency Attention Mechanism
  for TDNN and CNN-LSTM-TDNN, with Application to Language Identification},''
  in \emph{Proc. Interspeech 2019}, 2019, pp. 4080--4084. [Online]. Available:
  \url{http://dx.doi.org/10.21437/Interspeech.2019-1256}
\BIBentrySTDinterwordspacing

\bibitem{lstm1}
W.~{Cai}, D.~{Cai}, S.~{Huang}, and M.~{Li}, ``Utterance-level end-to-end
  language identification using attention-based cnn-blstm,'' in \emph{ICASSP
  2019 - 2019 IEEE International Conference on Acoustics, Speech and Signal
  Processing (ICASSP)}, 2019, pp. 5991--5995.

\bibitem{lstm2}
Y.~{Tang}, G.~{Ding}, J.~{Huang}, X.~{He}, and B.~{Zhou}, ``Deep speaker
  embedding learning with multi-level pooling for text-independent speaker
  verification,'' in \emph{ICASSP 2019 - 2019 IEEE International Conference on
  Acoustics, Speech and Signal Processing (ICASSP)}, 2019, pp. 6116--6120.

\bibitem{lstm3}
C.~{Chen}, S.~{Zhang}, C.~{Yeh}, J.~{Wang}, T.~{Wang}, and C.~{Huang},
  ``Speaker characterization using tdnn-lstm based speaker embedding,'' in
  \emph{ICASSP 2019 - 2019 IEEE International Conference on Acoustics, Speech
  and Signal Processing (ICASSP)}, 2019, pp. 6211--6215.

\bibitem{lstm4}
Y.~{Zhao}, T.~{Zhou}, Z.~{Chen}, and J.~{Wu}, ``Improving deep cnn networks
  with long temporal context for text-independent speaker verification,'' in
  \emph{ICASSP 2020 - 2020 IEEE International Conference on Acoustics, Speech
  and Signal Processing (ICASSP)}, 2020, pp. 6834--6838.

\bibitem{two}
\BIBentryALTinterwordspacing
L.~Song, Y.~Li, Z.~Jiang, Z.~Li, X.~Zhang, H.~Sun, J.~Sun, and N.~Zheng,
  ``Rethinking learnable tree filter for generic feature transform,'' in
  \emph{Advances in Neural Information Processing Systems}, H.~Larochelle,
  M.~Ranzato, R.~Hadsell, M.~F. Balcan, and H.~Lin, Eds., vol.~33.\hskip 1em
  plus 0.5em minus 0.4em\relax Curran Associates, Inc., 2020, pp. 3991--4002.
  [Online]. Available:
  \url{https://proceedings.neurips.cc/paper/2020/file/2952351097998ac1240cb2ab7333a3d2-Paper.pdf}
\BIBentrySTDinterwordspacing

\bibitem{dilated}
F.~Yu, V.~Koltun, and T.~Funkhouser, ``Dilated residual networks,'' in
  \emph{Proceedings of the IEEE conference on computer vision and pattern
  recognition}, 2017, pp. 472--480.

\bibitem{gao2019res2net}
S.~Gao, M.-M. Cheng, K.~Zhao, X.-Y. Zhang, M.-H. Yang, and P.~H. Torr,
  ``Res2net: A new multi-scale backbone architecture,'' \emph{IEEE transactions
  on pattern analysis and machine intelligence}, 2019.

\bibitem{attention}
A.~Vaswani, N.~Shazeer, N.~Parmar, J.~Uszkoreit, L.~Jones, A.~N. Gomez,
  L.~Kaiser, and I.~Polosukhin, ``Attention is all you need,'' \emph{arXiv
  preprint arXiv:1706.03762}, 2017.

\bibitem{hsresnet}
P.~Yuan, S.~Lin, C.~Cui, Y.~Du, R.~Guo, D.~He, E.~Ding, and S.~Han,
  ``Hs-resnet: Hierarchical-split block on convolutional neural network,''
  \emph{arXiv preprint arXiv:2010.07621}, 2020.

\bibitem{sparse}
G.~Zhao, J.~Lin, Z.~Zhang, X.~Ren, Q.~Su, and X.~Sun, ``Explicit sparse
  transformer: Concentrated attention through explicit selection,'' \emph{arXiv
  preprint arXiv:1912.11637}, 2019.

\bibitem{vox1}
A.~Nagrani, J.~S. Chung, and A.~Zisserman, ``Voxceleb: a large-scale speaker
  identification dataset,'' \emph{arXiv preprint arXiv:1706.08612}, 2017.

\bibitem{vox2}
\BIBentryALTinterwordspacing
J.~S. Chung, A.~Nagrani, and A.~Zisserman, ``Voxceleb2: Deep speaker
  recognition,'' \emph{CoRR}, vol. abs/1806.05622, 2018. [Online]. Available:
  \url{http://arxiv.org/abs/1806.05622}
\BIBentrySTDinterwordspacing

\bibitem{sitw}
M.~McLaren, L.~Ferrer, D.~Castán~Lavilla, and A.~Lawson, ``The speakers in the
  wild (sitw) speaker recognition database,'' 09 2016, pp. 818--822.

\end{thebibliography}


\end{document}